\documentstyle[twoside,fleqn,espcrc2,graphicx]{article}
\newcommand{\QQB}{$Q\bar{Q}$~}

\newcommand{\qqb}{$q\bar{q}$~}

\newcommand{\AmS}{{\protect\the\textfont2
  A\kern-.1667em\lower.5x\hbox{M}\kern-.125emS}}

\title{Review on string breaking\\
-- the query in quest of the evidence --}

\author{K. Schilling\address{von Neumann Institute of Computing, 
        c/o Research Center J\"ulich, and DESY Hamburg, \\ 
        D-52425 J\"ulich, Germany}%
        \thanks{on leave of absence from Physics Department, University of Wuppertal}
        }
       
\begin{document}
\begin{abstract}
  Considerable progress has been achieved recently in the observation of
  string breaking within non-Abelian Higgs models, by use of multi-channel
  methods allowing for broken string states. Similarly, in pure gauge theory
  this approach has been shown to reveal string breaking for color charges in
  the adjoint represenation.  For QCD with dynamical fermions, one needs
  substantial progress in noise reduction, however, to render such techniques
  viable. 

\end{abstract}
\maketitle 
\section{INTRODUCTION}
Twenty years after the demonstration of confinement in {\it quenched} QCD
simulations \cite{creutz} we have beautiful evidence of the formation of
colour flux tubes between static colour charges \cite{schlichter_internet}.
As to the verification of their fission, it has been anticipated ever since
that {\it full} QCD calculations should reveal string breaking (SB), in form
of a screening behaviour in the static potential $V(R)$.

\begin{figure}[htb]
\vspace{9pt}
\includegraphics[width=70mm]{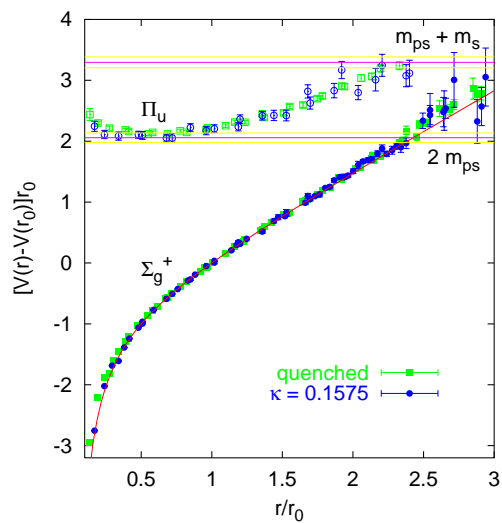}
\vspace{-10mm}
\caption{Update of full
  QCD potential from T$\chi$L (at $\beta = 5.6$) \cite{sesam_bali}, in
  comparison to their quenched result at $\beta_{q} = 6.0$, with 
Sommer radius $r_0$ \cite{sommer}.}
\label{fig:bbplot}
\end{figure}

While studies on Polyakov loop correlators in the temperature range $ .68 \leq
T/T_c \leq .94$ \cite{detar} reviewed at LATTICE98 \cite{kuti} did indeed
provide clear evidence for such phenomenon ({\it cf.} also the arguments
presented in \cite{gliozzi}), the head-on approach for detection of SB signals
from mere Wilson loop studies at $T = 0 $ has not proven successful
\cite{headon}.  This situation is illustrated in Fig.  \ref{fig:bbplot} {\it
  e.g.}, with updated T$\chi$L data \cite{sesam_bali}.

But why bother about SB if one expects it anyhow?  The motivations to focus on
the challenge are obvious: (a) the $R$-asymptotia of $V(R)$ is the most
obvious infrared quantity that we can target on the lattice and hence
important to be  under control, (b) SB can teach us both about the techniques
to handle mixing problems and hadron decays like
$\Upsilon  \rightarrow B\bar{B}$
  on the lattice \cite{oxford}. Lastly,
understanding hadronization in the static interaction scenario of full QCD
will certainly help to further substantiate previous quenched results on
confinement \cite{bali_schilling}.  For we should remember that transfer
matrix studies yield at best numerical estimates of
{\it upper bounds} to $V$ as long as we are analyzing asymptotic (in Euclidean
time $T$) behaviour of Wilson loops $W$ in terms of $dln W(R,T)/dT  \rightarrow
-V(R)$, at finite $T$.

\section{VISIBILITY:  5 LATTICE SPACINGS}
Quenched analyses of static potentials rely -- apart from high statistics --
on the combined action of three signal enhancing techniques : (a) smearing the
spatial links in order to suppress excited state contaminations, (b) analytic
multihit noise reduction on the  time-like links, (c) exploiting
loop  averaging effects on each configuration by shifting and measuring $W$ all
over the entire lattice. 

\begin{figure}[t]
\vspace{9pt}
\includegraphics[width=70mm]{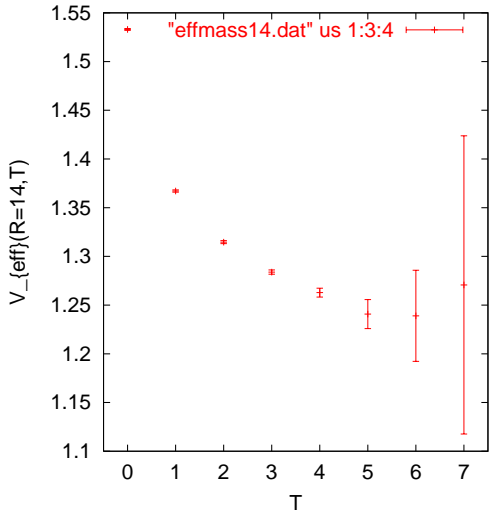}
\vspace{-10mm}
\caption{Plateau formation in $V_{eff}(2.4 r_0)$  at
  $\beta = 5.6$ and $\kappa_{sea} = .1575$, from T$\chi$L configs. \cite{sesam_bali}}
\label{fig:effmass}
\end{figure}

In full QCD we are in lack of high statistics samples of vacuum configurations
(for cost reasons) and moreover, we have to abandon multihitting as it becomes
unfeasable in presence of long range quark loop effects.  In
Fig.\ref{fig:effmass} I demonstrate the quality of the resulting plateau in
the effective potential, $V_{eff}(2.4 r_0,T)$.  Evidently, one runs into rapidly
increasing errors when going beyond $T = 5-6$!

The lesson to be learnt from Fig.\ref{fig:effmass} is obviously that -- in
order to uncover SB -- one better aims at achieving an improved overlap with
the true ground state \cite{guesken}, in terms of an earlier onset of the
plateau, in a region where the resolution of the effective potential still
suffices,
at time separations $3, 4,$ and $5$. 

\begin{figure}[t]
\vspace{9pt}
\includegraphics[width=70mm]{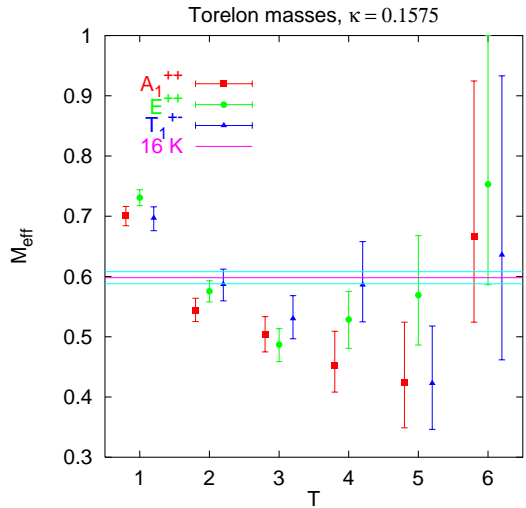}
\vspace{-10mm}
\caption{Effective torrelon masses from SESAM configurations at $\kappa_{sea}
  = .1575$ \cite{bali_torelon}.}
\label{fig:torelons}
\end{figure}

Let me just mention in passing an alternative strategy to establish SB: (a)
determine in full QCD the masses of torelons, i.e.  the spatial wrappers (of
length $L$) around the lattice \cite{michael_torelons}; (b) compute an
effective string tension, $K$, like in quenched QCD from a fit to the
potential at $r < 2 r_0$ where string breaking is not yet expected to set in;
(c) torelon effective masses undershooting the string energy would provide
evidence for SB.  As illustrated in Fig.  \ref{fig:torelons} the SESAM data on
the $16^3\times 32$ lattice at their smallest quark mass exhibits {\it some
  weak evidence for SB}, by crossing the level $\sigma L = 16 K$, though at
rather high noise level.

Another aspect which has been brought forward recently \cite{trottier}
is to exploit coarse grained spatial lattices by help of improved actions.

\section{MULTICHANNEL APPROACH}
The obvious strategy for improving the ground state overlap is to enlarge the
operator space to a multichannel approach {\it comprising
\begin{figure}[b]
\hspace{20mm} \includegraphics[width=30mm]{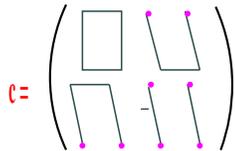}
\vspace{-5mm}
\caption{The two channel transfer  matrix $C(T)$ in the 
  non-Abelian Higgs model, with due tribute to Pisa: the bullets stand for
  scalar mass insertions.}
\label{fig:matrix}
\end{figure}
hadronization}. This would be very much in the spirit of multi-operator
variational ans\"atze as devised and used largely in the context of glueball
studies.  In the present context strong coupling methods have been used to
demonstrate \cite{strong_coupling} that inclusion of the \QQB \qqb channel in
addition to the static quark-antiquark \QQB (with fluxtube in between) looks
like a promising path for uncovering SB.  It implies however the use of large
loops with color partly transmitted by insertion of light quark propagators
(instead of by gauge links only).
\begin{figure}[t]
\vspace{-3mm}
 \includegraphics[width=55mm]{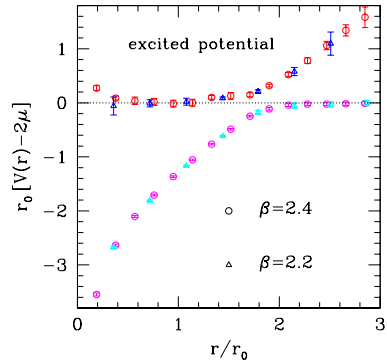}
\vspace{-10mm}
\caption{Scaling of ground and excited states potential
in the SU(2) Higgs model
  \cite{knechtli_latt99}.}
\label{fig:scaling}
\end{figure} 
This feature precludes us from applying easy loop averaging  over
the entire lattice, since light quark propagators are too expensive to compute
for all source points on the lattice.
\section{SB WITHOUT FERMIONS}
Given the problems to deal with fermions refs.  \cite{knechtli1,wittig1}
followed an old suggestion of Bock et al \cite{bock} to study the mechanism of
SB in an easier setting, namely by resorting to non-Abelian Higgs fields.

In the two-channel scenario (see Fig. \ref{fig:matrix}) one is faced with a
generalized eigenvalue problem of the transfer matrix, 
for evolution from $T_0 =0$ to $T$ , with eigenvalues $\lambda_i(T,T_0)$
\cite{wolff}
\begin{equation} 
 C(T) \; \vec{c}_i  = \lambda _i(T, T_0)\quad  C(T_0)\; \vec{c}_i ,
\end{equation}
the index labeling ground ($i=0$) and excited state ($i=1$).  In a
contribution to this conference Knechtli et al \cite{knechtli_latt99} have
presented a high statistics scaling study of an SU(2) Higgs model in 3+1
dimensions, with matter field $\Phi$ in the fundamental representation, in the
confining phase.  As can be seen from Fig. \ref{fig:scaling} they can neatly
trace both ground and excited states in a scaling regime, with a clear-cut gap
in between.

\begin{figure}[t]
 \includegraphics[width=55mm]{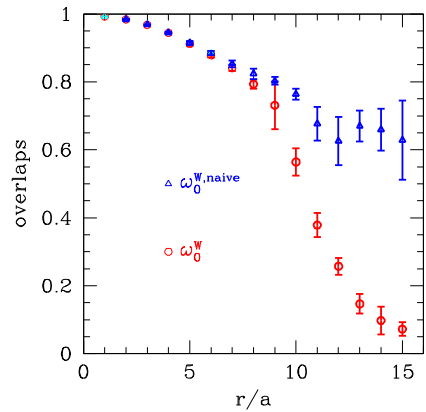}
\vspace{-10mm}
\caption{The overlap of the Wilson loop operator, $\omega_0$, with the broken 
  string ground state \cite{knechtli_latt99}.}
\label{fig:overlap}
\end{figure}

Moreover they show that the overlap function of the Wilson loop operator to
the true ground state, $\omega_0^W(R)$, computed according to
\begin{equation}
\Omega (T) = (\vec{c}_0, C(T) \; \vec{c}_0) = \sum_n  \omega _n \exp{(-V_n T)}, \end{equation}
with $\Omega(0) = 1$ exhibits a steeply falling $R$-dependence, 
as shown in Fig.  \ref{fig:overlap}.  They illustrate moreover
that a naive one-channel analysis is  
manifestly misleading, as it yields   an 
overlap estimate,  $ \omega^{W,naive}_0$, that fakes 
a good  projection  of the large Wilson loops to the ground state!!

\begin{figure}[h]
\vspace{-25mm}
 \includegraphics[width=80mm]{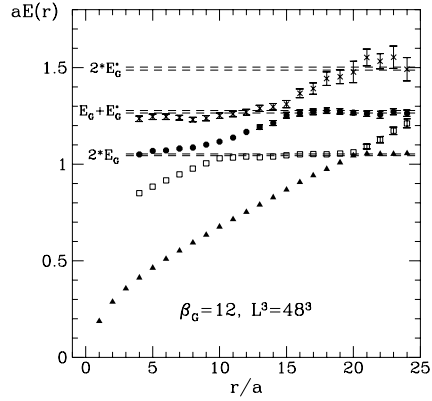}
\vspace{-15mm}
\caption{SB observed for three different excitation levels of the adjoint
string, from ref.
  \cite{wittig_adj}.}
\label{fig:wittig_adj}
\end{figure}

As yet another check for the efficiency of the two-channel approach in
revealing SB, two detailed studies \cite{wittig_adj,stephenson} were addressed
recently to the adjoint flux tube formation between static color charges in
the adjoint representation of SU(2) gauge theory  in 2+1
dimensions. Such strings are not protected from breaking up, even in absence
of fermions \cite{michael_adj}.  In the correlation matrix $C$ (see Fig
\ref{fig:matrix}) the mass insertions are now to be replaced by gluelump
operators\cite{michael_glue}, $ \mbox{gluelump}_a^x = Tr\big(\mbox{clover}_a^x
\sigma_a\big)$ (with color $a$ and site $x$) to be linked by the color
flux operator $Tr\big( \sigma^a \; U_{x\rightarrow y} \; \sigma^b
\;U^\dagger_{x\rightarrow y}\big) $.  Fig.  \ref{fig:wittig_adj} highlights
beautifully the efficiency of the multichannel ansatz in revealing, in
addition to the ground state, another three excited states, along with
resolution of their level crossings!
\begin{figure}[t]
\vspace{-60mm}
\hspace{-60mm}
\includegraphics[width=100mm,angle=270]{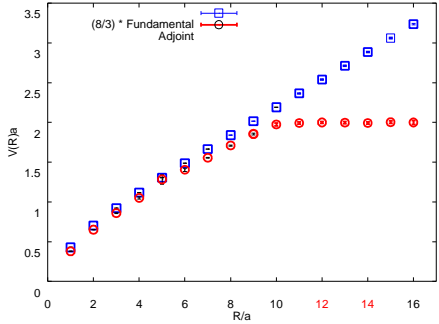}\\
\caption{SB in adjoint potential compared to confinement in
static potential in fundamental representation
in 2+1d SU(2) g.t.,  from ref.  \cite{stephenson}.}
\label{fig:bothpotential_steph}
\end{figure}

Having convinced ourselves that an operator set extended to include physically
motivated channel mixing is the right way to go, let us return to our original
problem, i.e.  QCD with fermions.
\begin{figure}[t]
\includegraphics[width=60mm,angle=0]{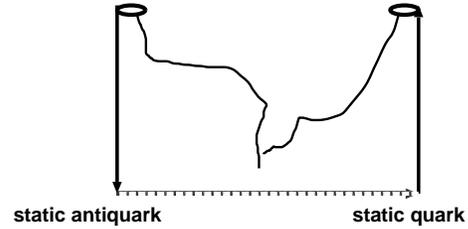}
\vspace{-5mm}
\caption{Transition matrix element from state \QQB to state \QQB \qqb,
with source/sink  smearing.}
\label{fig:transition}
\end{figure}
\begin{figure}[h]
\vspace{-80mm} 
\includegraphics[width=130mm,angle=270]{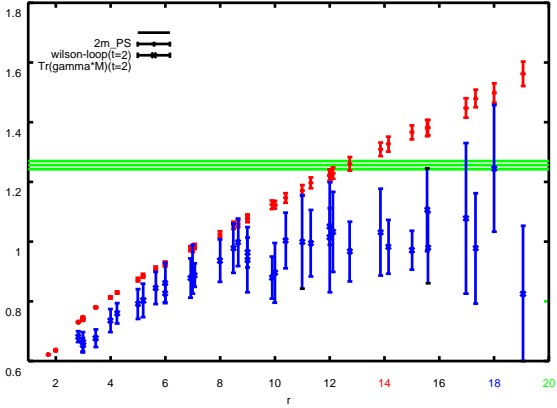}
\caption{Noise on local masses {\it vs} $R$, at $T = 2$: small (large) error
  points refer to Wilson loops (transition matrix elements,
  $\langle2|C|1\rangle$, {\it cf.} Fig. \ref{fig:transition}); 
 horizontal lines mark the error band for $2 m_{PS}$.
 Analysis done
  on T$\chi$L configs, at $\kappa_{sea} = .1575 $ 
and $r_0 \approx 5.9a$ \cite{struckmann}.}
\label{fig:struckmann}
\end{figure}

\section{UNCOVER SB IN QCD?}
If we wish to adopt the multichannel strategy to the QCD situation, we have to
deal with loops containing insertion of light quark propagators, like in the
transition correlator $\langle2|C|1\rangle$ depicted in Fig.
\ref{fig:transition}.  

Fig.  \ref{fig:struckmann} \cite{struckmann} illustrates the noise situation
with the example of local masses from the transition correlator,
$\langle2|C|1\rangle$.  While source smearing helps a lot, doing without loop
averaging makes things hopeless, even at a small value of $T$, such as $T = 2
(!)$.  This can best be seen by comparing to the noise level in the
corresponding Wilson loop, $\langle1|C|1\rangle$!  One should be aware that
the noise situation is considerably aggravated for $\langle2|C|2\rangle$, at
$T = 3$ or $4$! Similar observations have been reported by Lacock for the MILC
collaboration \cite{lacock}.

In order to get noise under control one obviously needs the computation of
light quark propagators $P(x\rightarrow y) = M^{-1}_{xy}$ on source locations
sampled over a large region, $R$, of the lattice volume. A promising program
in this direction has been launched recently by Michael et
al.\cite{michael_penn} Their proposal is to start out from a Gaussian estimate
for inverse Dirac operator in  terms of a random scalar field $\Phi$.
\begin{equation}
 M^{-1}_{xy}  
= \frac{1}{Z}\int D\Phi \quad
\Phi_y^*\Phi_x\exp{\{-\Phi^*M\Phi/2\}} \quad .
\end{equation}
The estimator introduces additional noise whose variance can be minimized by a
multihit (analytical integration) on $\Phi$ over the interior region of $R$,
$I = R - \partial R$. In praxi, this requires  computation of  an inverse
block matrix $\bar{M}^{-1}$ (by iterative solver) after each update of the
$\Phi$-field on the boundary $\partial R$, as borne out in the replacement:
 \begin{eqnarray}
 \Phi_x \rightarrow v_x 
&=& \int_{\Phi \subset I } D\Phi \quad \Phi_z
\exp{\{-\Phi^*M\Phi/2\}} \nonumber \\  &=& 
- \bar{M}^{-1}_{xx'}\tilde{M}_{x'z}\Phi_z \quad .
\end{eqnarray}
Here $\bar{M}$ ($\tilde{M}$) is a block matrix with support $\subset I = R -
\partial R$ ($R$) and $z \in \partial R$ .

So far this noise reduction technique has been applied successfully to the
determination of forces between two static-light mesons \cite{michael_hl}. It
remains to be seen whether it can bring decisive progress in the treatment of
QCD string fission.

\section{PISA EST OMEN?}
While in confining models without fermions, SB has been clearly seen at work,
we have at this stage no real compelling evidence for string breaking from
simulation of QCD with dynamical fermions.

\begin{figure}[htb]
\vspace{9pt}
\includegraphics[width=45mm]{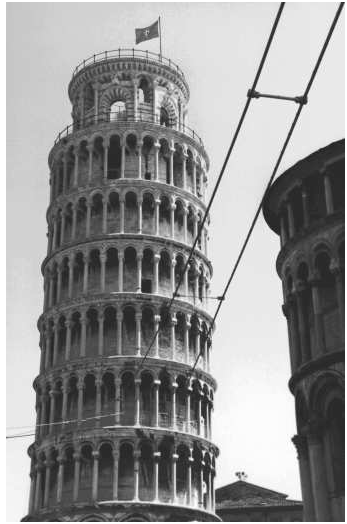}
\caption{The Pisa  model for  b-confinement.}
\label{fig:tower}
\end{figure}

In any case, from meditating on Fig.  \ref{fig:tower} which is a zoom to this
years conference poster a superstitious person might suppose that at present
Pisa would not offer favourable auspices for the occurence of string breaking.
So for the time being, let me finish by quoting a famous last century tourist
who felt frustration after climbing (the tower?): ``{\it Spitze Steine --
  M\"ude Beine -- Aussicht keine -- Heinrich Heine''.} What about   Heine
being wrong? 

\vspace{5mm}
{\bf Acknowledgements}

My great thanks to the organizers for the superb conference.  I enjoyed
interesting discussions with G. Bali, M. Peardon, and T.  Struckmann during
the preparation of this talk. I am grateful to G. Bali for his help in
improving the manuscript.

\end{document}